\documentclass[twocolumn,twocolappendix]{aastex701}

\usepackage{caption}

\newcommand{\msun}{M$_{\sun}${}}
\shorttitle{A dwarf galaxy group in the local universe}
\shortauthors{Paudel et al.}

\begin{document}

\title{Different Origins of Nucleated and Non-nucleated Dwarf Elliptical Galaxies:\\
Identified by the Deep-learning 
 }

\author[orcid=0000-0003-2922-6866,gname='Sanjaya',sname='Paudel']{Sanjaya Paudel}
\affiliation{Department of Astronomy \& Center for Galaxy Evolution Research, Yonsei University, Seoul 03722, Republic Of Korea }
\affiliation{Nepal Astronomical Society, Kathmandu, Nepal}
\email{sanjpaudel@google.com}  

\author[orcid=0000-0002-5513-5303,gname='Cristiano G.', sname='Sabiu']{Cristiano G. Sabiu} 
\affiliation{Natural Science Research Institute (NSRI), University of Seoul, Seoul 02504, Republic of Korea}
\email{csabiu@google.com}

\author[orcid=0000-0002-1842-4325,sname='Yoon']{Suk-Jin Yoon}
\affiliation{Department of Astronomy \& Center for Galaxy Evolution Research, Yonsei University, Seoul 03722, Republic Of Korea }
\email[show]{sjyoon0691@yonsei.ac.kr}

\author[orcid=0000-0002-8040-6902,sname='Chhatkuli']{Daya Nidhi Chhatkuli}
\affiliation{Department of Physics, Tri-Chandra Multiple Campus, Tribhuvan University, Kathmandu, Nepal}
\email{chhatkulidn@gmail.com}

\author[orcid=0000-0003-0960-687X,sname='Zee']{Woong-Bae G. Zee  }
\affiliation{School of Liberal Studies, Sejong University, 209 Neungdong-ro, Gwangjin-gu, Seoul 05006, Republic of Korea.}
\email{galaxy.wb.zi@gmail.com}

\author[orcid=0000-0002-6841-8329,sname='Yoo']{Jaewon Yoo  }
\affiliation{Korea Astronomy and Space Science Institute (KASI), Daedeokdae-ro, Daejeon 34055, Republic of Korea}
\email{jwyoo@kias.re.kr}

\author[orcid=0000-0002-2799-9112,sname='Adhikari']{Binod Adhikari }
\affiliation{Department of Physics, St. Xavier’s College, Tribhuvan University, Kathmandu, Nepal}
\email{sjyoon0691@yonsei.ac.kr}

\begin{abstract}
Dwarf elliptical galaxies (dEs) are the dominant population in galaxy clusters and serve as ideal probes for studying the environmental impact on galactic evolution. A substantial fraction of dEs are known to harbor central nuclei, which are among the densest stellar systems in the Universe. The large-scale distribution and the underlying origin of nucleated and non-nucleated dEs remain unresolved. Using a state-of-the-art machine learning framework, we systematically scan the Virgo cluster region ($15\arcdeg \times 20\arcdeg$ centered at $R.A. = 187.2\arcdeg$ and $Dec. = 9.6\arcdeg$) and construct the largest homogeneous sample of dEs (of total 2,123) with robust nucleus classifications.  We find that nucleated dEs are more spatially clustered and exhibit a stronger association with massive galaxies than their non-nucleated counterparts. This suggests that most nucleated dEs likely formed alongside massive galaxies within the cluster (i.e, the {\it in-situ} formation). In contrast, non-nucleated dEs are more widely distributed across the cluster and align more closely with Virgo's global potential well, as traced by the cluster's hot gas. This indicates that most non-nucleated dEs originated outside the cluster (i.e, the {\it ex-situ} formation) and were later accreted and redistributed within it. Our findings shed new light on how dEs and their central nuclei form and evolve.
\end{abstract}


\section{Introduction}

Dwarf galaxies are by far the most numerous in the Universe and exist in all kinds of environments, from dense galaxy clusters to cosmic voids \citep{Binggeli85,Rekola05,Makarov17}. 
Dwarf elliptical galaxies (dEs) among dwarfs, mainly known for their old stellar population and low surface brightness, are a dominant population, outnumbering all other galaxy types in dense environments \citep{Ferguson94,Boselli08,Kormendy09}. 
Due to their shallow gravitational potential, they are more susceptible to environmental effects than giant galaxies \citep{Boselli08}.
Their intrinsically low surface brightness has significantly challenged their detection and classification. In recent years, studies of small to intermediate-sized samples of dEs using various approaches \citep{Ordene18,Habas20,Paudel23} have consistently shown that dEs exhibit an extreme form of the morphology--density relation--namely, they are the most prevalent galaxy type in dense cluster environments, yet rare in low-density regions such as fields and voids \citep{Geha12,Lieder13,Paudel23}.

Two primary theories explain the origin of dEs in cluster environments: the \textit{`nature'} hypothesis, which suggests that dEs are ancient, primordial galaxies formed early in the Universe, and the \textit{`nurture'} hypothesis, which proposes that their evolution is predominantly shaped by environmental interactions \citep{Boselli08,Kormendy09,Annibali11}.

Recent studies, however, indicate a more complex scenario involving preprocessing, where dEs undergo environmental influences in smaller groups before being accreted into larger clusters \citep{Fujita04,Cortese06,Su21,Lokas23}. This scenario predicts a close spatial alignment between the distributions of dEs and massive galaxies, as they are often accreted together. However, the limited sample size of dEs has hindered a detailed investigation of their relationship with massive galaxies. Moreover, dEs exhibit diverse morphologies, including the presence or absence of central nuclest star cluster (NSC) \citep{Cote06,Lisker07,Paudel11,Urich17}. NSCs are compact systems (half light radii = 1–50 pc) with stellar masses of $10^{4}$–$10^{8}$ \msun, typically larger and denser than globular clusters (GCs), with central densities similar to the densest GCs and ultra-compact dwarfs (UCDs) \citep{Lauer98,Hilker99,Drinkwater00}.

The conditions leading to the formation of nuclei and the factors that determine why some dEs possess nuclei while others do not are still not fully understood. 
Although nucleation is more common in brighter dEs, observational biases that make them easier to detect may have influenced this finding \citep{Ordenes18,Janssen19,Poulain21}.
To accurately assess the prevalence and physical origin of nucleated dEs, a statistically significant sample of faint dEs is crucial.

NSCs are believed to form through two primary mechanisms: (i) the dissipationless infall and merger of dense star clusters driven by dynamical friction, and (ii) dissipative in-situ star formation triggered by gas inflows \citep{Tremaine75,Bekki06}.  However, recent observational and theoretical studies suggest that globular cluster mergers play a dominant role in dwarf galaxies. For instance, \citet{Neumayer20} reviewed several arguments supporting this scenario, while \citet{Fahrion21} reported metal-poor nuclear clusters indicative of merged globulars. \citet{Poulain25} have directly observed a globular cluster merger forming a nuclear cluster. Environmental trends further support this link: \citet{Carlsten22} find that the occupation fraction of globular clusters closely follows that of nuclear clusters in dwarfs. On the other hand, hybrid pathways, such as the coalescence of gas-rich clusters, have also been proposed \citep{Agarwal11,Guillard16,Paudel20}. Observations reveal that NSCs host complex stellar populations, often younger and more metal-rich than typical globular clusters, consistent with contributions from multiple formation channels \citep{Rossa06,Walcher05,Paudel11}. 

The occupation fraction of NSCs shows a clear dependence on environment, with galaxies in denser regions—such as group and cluster cores—exhibiting higher nucleation rates (e.g. \citealt{Sanchez19, Hoyer21, Paudel23, Zanatta24}). This environmental trend likely reflects both enhanced external interactions and differences in internal evolutionary pathways. Frequent tidal encounters and minor mergers in dense environments may promote the inward migration of globular clusters through dynamical friction, while gas-rich interactions can channel material toward galaxy centers, triggering in-situ star formation \citep{Dolcetta08,Sedda14,Schiavi21,Roman23}. In contrast, isolated or low-density environments may suppress these processes, leading to a lower probability of NSC formation. Nevertheless, the relative contribution of these mechanisms—and the physical reasons why some galaxies remain devoid of NSCs—remain key open questions in understanding the co-evolution of galaxies and their nuclei.

The Virgo cluster, as the nearest massive galaxy cluster, provides a unique opportunity to study the detailed environmental effects on galaxy evolution, with a particular focus on its dominant population of dEs. 
Located 16.5 Mpc away \citep{Mei07}, the large-scale automated identification of dEs within the cluster has long been challenging due to their low surface brightness.
To address this challenge, we have developed a deep learning framework capable of semi-automatically detecting dEs at the distance of the Virgo cluster, scanning every pixel of sky.

\begin{figure*}
\includegraphics[width=12.2cm]{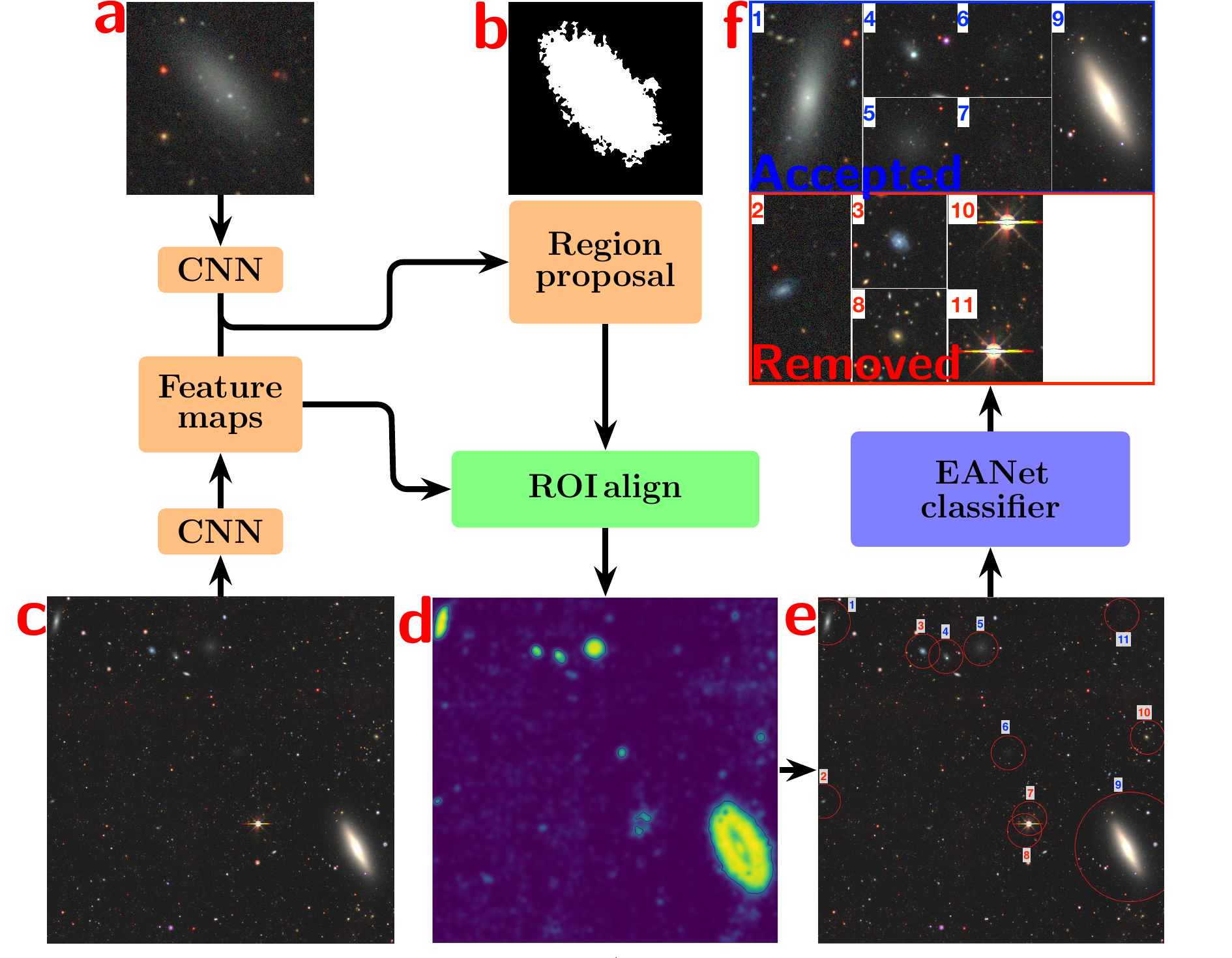}
\caption{
The neural network architecture of our deep learning framework for dE galaxy detection and localization. 
Our framework extends the  R-CNN to predict both segmentation masks and class probabilities. 
(a) The input training RGB image is processed by a custom layer CNN to produce multi-scale feature maps. 
(b) A region proposal network (RPN) identifies candidate regions likely to contain galaxies. 
(c) An example of the input inference image. 
(d) The probability map produced after processing through CNN and ROI alignment. 
(e) The potential dE candidates are identified. 
(f) An External Attention Network classifier, trained on a dataset of visually identified dE, assigns each region to a class (dE or non-dE).
The final output shows ``Accepted" (blue) and ``Removed" (red) regions, effectively distinguishing true detections from false positives.}
\label{schm}
\end{figure*}

\section{Detection}

Traditional methods for identifying such sources typically rely on pre-built catalogs generated by automated algorithms, such as Source Extractor, followed by applying criteria to select objects of interest.
In contrast, our approach directly identifies dEs by leveraging expert knowledge gained through the visual analysis of previously identified dEs. 
This methodology circumvents the pre-selection biases inherent in automated detection software.
To further enhance and scale this manual process, we incorporate state-of-the-art machine learning techniques. 
The models scan every pixel of the sky of the input images, ensuring complete coverage beyond the limitations of human visual inspection. 

\subsection{Training Sample}
In our earlier study, we constructed a comprehensive catalog of dEs within the nearby Universe ($z \leq 0.01$) through systematic visual inspection of the imaging data from the Legacy Survey \citep{Paudel23}. The resulting catalog contains more than 5,000 dEs that span a wide range of environments, including clusters, groups, and fields. Of these, 1,324 galaxies had previously published spectroscopic redshifts. To improve the spectroscopic completeness of the sample, we cross-matched our catalog with the first data release of the Dark Energy Spectroscopic Instrument \citep[DESI DR1][]{Desidr1}. This effort yielded redshift confirmations for an additional 936 dEs, among which 150 had not been previously identified in the literature. The expanded catalog, therefore, consists of 5,154 dEs in total, with spectroscopic redshifts available for 2,110 objects.

\subsection{Model Architecture}

We developed a detection pipeline that integrates a region-based convolutional neural network (R-CNN) with an External Attention Network (EANet; \citealt{Guo21}) to identify and classify dwarf galaxies. This approach is particularly effective for low-surface-brightness systems, which often have diffuse morphologies and are difficult to separate from background fluctuations or foreground contaminants. The R-CNN employs a region proposal network (RPN) that extracts multi-scale features from a deep backbone (e.g., ResNet; \citealt{He16}) to localize candidate galaxies. These candidates are then refined through Region of Interest (ROI) alignment, and the resulting features are classified with EANet, which embeds image patches into feature vectors and applies an efficient external attention mechanism to capture global context with linear complexity.

Figure \ref{schm} illustrates the workflow of our detection pipeline. The process begins with a three-channel (red, green, and blue) 256\,$\times$\,256 pixel input image (panel $a$). 
This image is passed through an 8-layer CNN to generate feature maps, capturing multi-scale representations of potential galaxy structures. 
The CNN, comprising $\sim$400,000 trainable weights, outputs a 64\,$\times$\,64 feature map highlighting regions likely associated with dEs.
Next, a region proposal network (RPN, panel $b$) processes the feature map to identify candidate regions that are most likely to contain dE galaxies. 
These candidate regions are further refined through Region of Interest (ROI) alignment (panels $c$, $d$, and $e$), ensuring precise localization by adjusting the feature maps. Finally, the aligned features are classified using an EANet (panel $f$). 

Following automated classification, we visually examined the confirmed dE candidates to ensure the reliability of the results, forming our final catalog. 
During this process, we observed some false positives, where the model misclassified features such as the outskirts of large galaxies or tidal structures as dEs.

\begin{figure}
\includegraphics[width=\hsize]{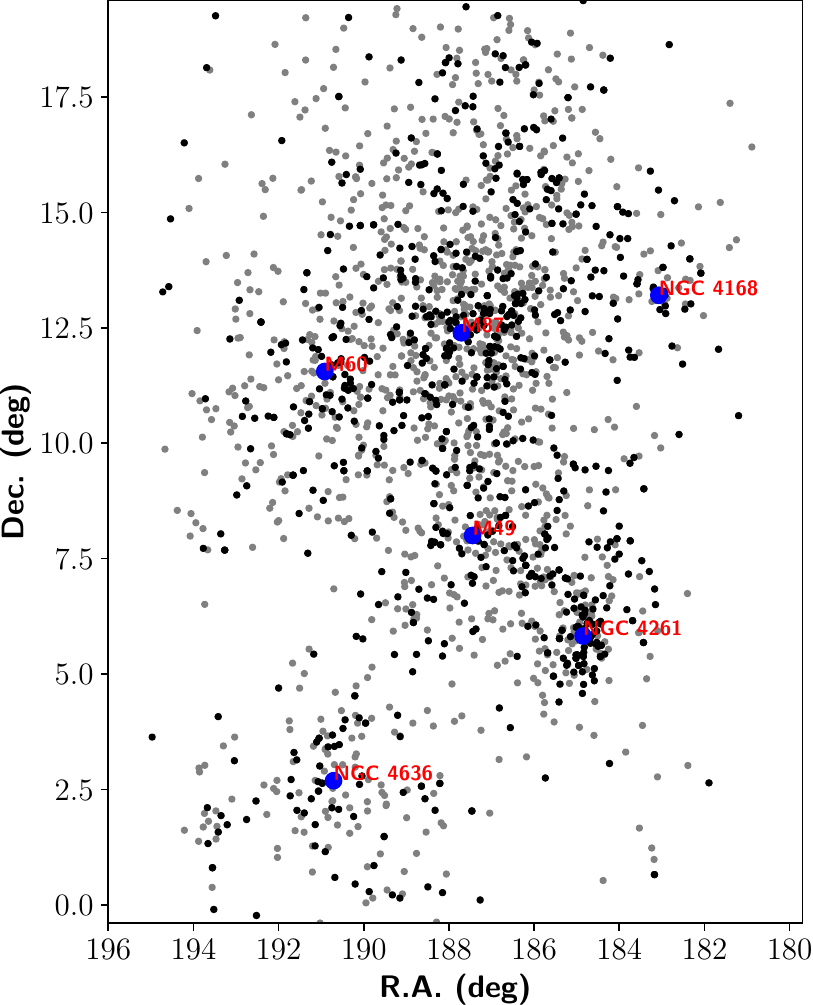}
\caption{
All sky distribution of identified dEs in our sample. A subset of the full sample which have a line-of-sight radial velocity measurement is shown in black.  We also highlight the six main subgroups of galaxies in the Virgo cluster area with large blue dot symbols and their names.}
\label{de_dist}
\end{figure}

\begin{figure}
\includegraphics[width=\hsize]{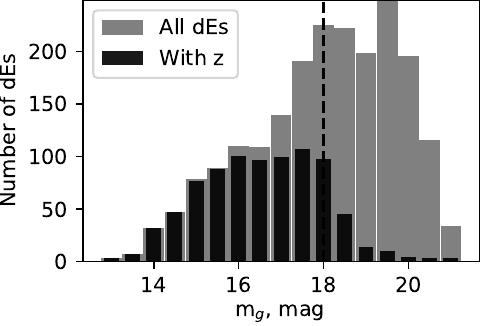}
\caption{
The $g$-band magnitude distributions of our dE sample (gray) and a subset having radial velocity information (black). }
\label{maghist}
\end{figure}

\section{Identified Sample}

We identified 2,123 dEs with $m_g < 20$ mag within the Virgo cluster region ($15\arcdeg \times 20\arcdeg$ centered at $R.A. = 187.2\arcdeg$ and $Dec. = 9.6\arcdeg$). Their sky positions are shown in Figure \ref{de_dist}, where black dots indicate dEs with confirmed redshifts ($z < 0.01$) compiled from the SDSS and DESI surveys. Figure \ref{maghist} presents the $g$-band magnitude distribution of our sample, peaking near 19.5 mag, with a median of 18.0 mag. This indicates that most of our dEs are fainter than the spectroscopic selection limit of large surveys such as the SDSS ($m_g = 18.0$ mag, marked by a vertical dashed line). The distribution of the spectroscopic subsample (black histogram) predominantly overlaps with the brightest portion of the sample, as expected.

The availability of an extensive catalog of Virgo cluster dwarf galaxies enables us to calibrate our semi-automated detection method. We assess its performance by comparing our detections with the well-characterized NGVS catalog \citep{Ferrarese16}, which provides a complete sample of Virgo cluster dwarfs within the central 4 deg$^{2}$ (2 $\times$ 2 deg) region. Our method successfully recovers 81\% of dwarf ellipticals (dEs) with apparent magnitudes m${g}$ $<$ 20 mag, and nearly 100\% of the brighter systems with m${g}$ $<$ 17 mag.

\begin{figure}
\includegraphics[width=\hsize]{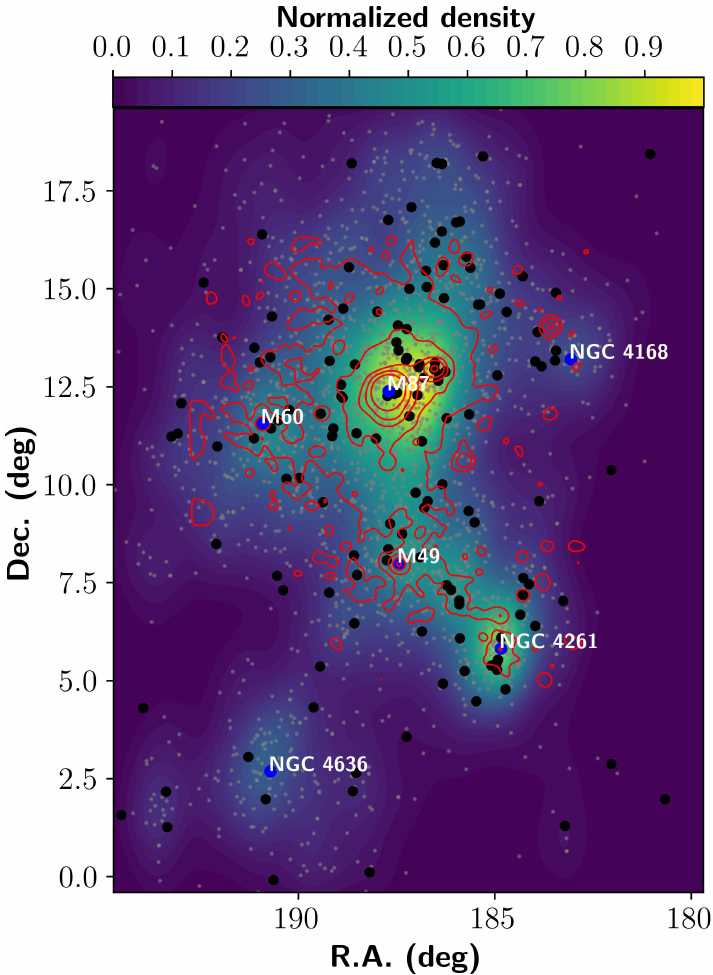}
\caption{
The surface number density map of dEs is compared with the spatial distribution of massive galaxies and X-ray emission in the Virgo cluster core region.
The surface number density map is constructed using 2D Gaussian Kernel Density Estimation (KDE) with a kernel width of 0\rlap{.}$^{\circ}$175 (50 kpc). 
The color bar at the top represents the normalized density.
Red contours indicate the surface flux density of hard X-ray emission (0.4–2.4 keV) in the Virgo cluster region, based on the ROSAT all-sky survey \citep{Bohringer94}.
Small gray dots denote individual dEs (from which the surface number density map is generated), while large black dots represent massive galaxies (M$_{*}$ $>$ 10$^{10}$ M$_{\odot}$).
Large blue dots with their names are the six main subgroups of galaxies.}
\label{de_gas_dist}
\end{figure}

\section{Results and Discussion}

In Figure \ref{de_gas_dist}, we construct a homogeneous surface number density map of dEs across the entire Virgo cluster area.  The resulting map reveals a diffuse, large-scale distribution of dEs extending approximately 6 degrees from the Virgo center, associated with the primary overdensity regions at galaxies like M87, M49, M60, NGC\,4168, NGC\,4261 and NGC\,4636. This distribution is significantly more diffuse and extended than that of massive galaxies (M$_{*}$ $>$ 10$^{10}$ M$_{\odot}$ ), marked by black dots. The dE density peak lies between M87 and M86, the two most massive members of the Virgo.  Notably, a substantial fraction of dEs inhabit void regions devoid of large galaxies within a 200 kpc radius.

\begin{figure}
\includegraphics[width=\hsize]{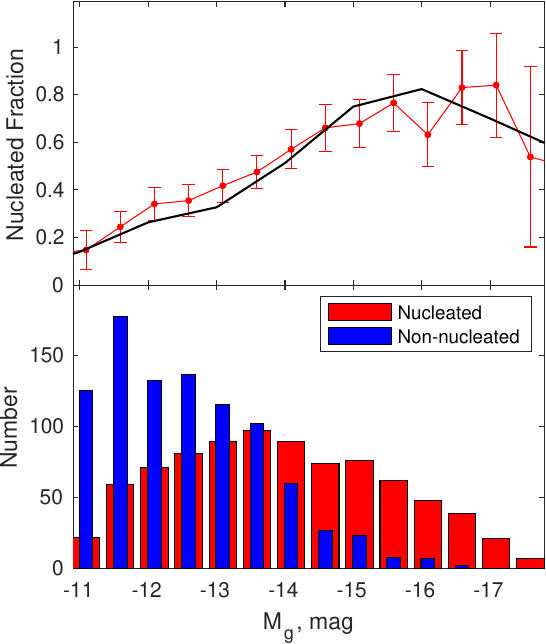}
\caption{Top: Fraction of nucleated dEs as a function of magnitude. The black line shows the nucleated fraction measured by \citet{Sanchez19} in the Virgo cluster core, while the red line represents our measurement for the full dE sample. The error bar represents the fraction of dEs that we were not able to classify into nucleated and non-nucleated. Bottom: $g$-band magnitude distributions of nucleated (red histogram) and non-nucleated (blue histogram) dEs, which we have calculated assuming all dEs are located at the average distance of the Virgo cluster, i.e., 16.5 Mpc.}
\label{nufrac}
\end{figure}

\begin{figure*}
\includegraphics[width=\hsize]{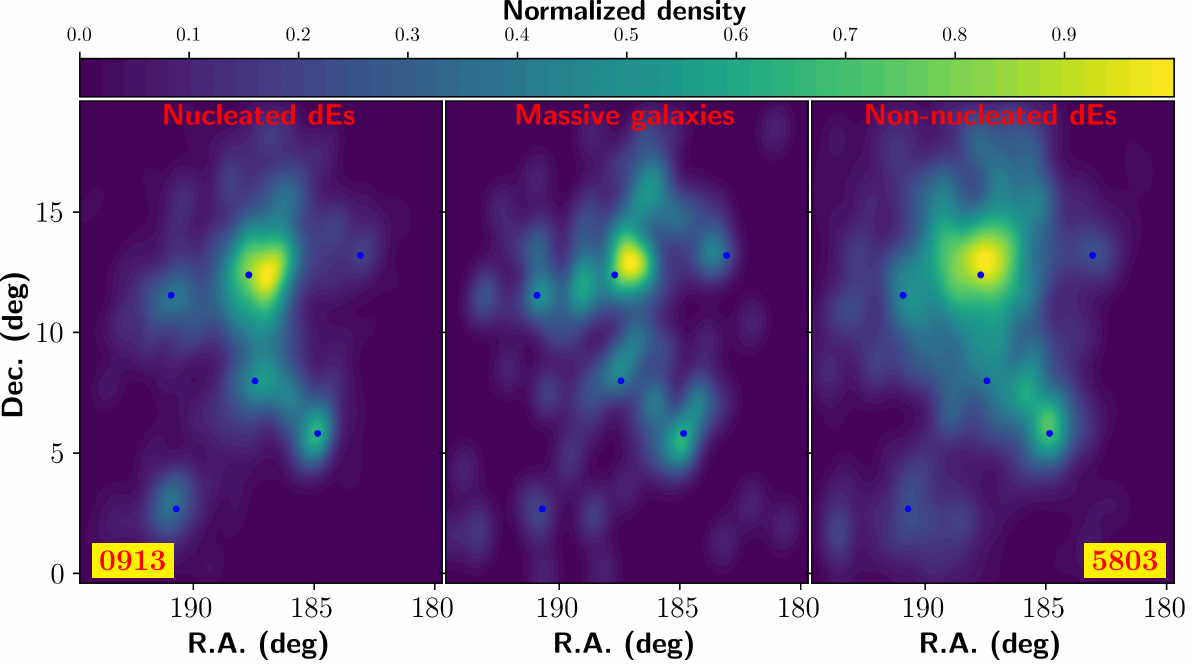}
\caption{
The galaxy distribution maps are generated using 2D Gaussian Kernel Density Estimation (KDE), following the same method as in Figure \ref{de_gas_dist}. We give the K-L divergence score for both nucleated and non-nucleated dEs at the bottom of each panel, comparing with that of massive galaxies.}
\label{nuclass}
\end{figure*}

By analyzing the hard energy band (0.4–2.4 keV) X-ray emission detected in the ROSAT All-Sky Survey \citep{Bohringer94}, we find that the inner contours, which represent regions of higher X-ray surface brightness, closely align with the positions of massive galaxies. In contrast, the outer contours, which trace the diffuse X-ray emission of the hot intracluster medium, show significant overlap with dEs. This suggests that dEs generally follow the large-scale global gravitational potential of the cluster, while massive galaxies create localized gravitational minima. 
Notably, this implies that at least a fraction of present-day cluster dEs---particularly those beyond the gravitational influence of large galaxies---were likely accreted in a more isotropic manner as the cluster evolved.

We manually classified dEs as nucleated or non-nucleated following \citet{Paudel23}. A nucleus is defined as a compact, PSF-sized point source at the galaxy center, representing a luminosity excess above the smooth stellar distribution \citep{Sanchez19,Paudel20}. Galaxies with central star formation or exceptionally high central surface brightness were excluded, as these features hinder reliable nucleus detection \citep{Lisker06,Urich17,Paudel20}, which include 13\% of total dEs.

Out of the full sample, we were able to classify 1,845 dEs into nucleated and non-nucleated systems. Among these, 851 were identified as nucleated, corresponding to a nucleated fraction of 46\%. Figure \ref{nufrac} (bottom panel) illustrates the $g$-band magnitude distributions of both classes. It reveals a clear luminosity dependence: nuclei are preferentially found in the brighter dEs, while the faintest systems are more likely to be non-nucleated. This is evident from the red histogram, which dominates the bright end of the distribution, whereas the blue histogram increasingly dominates at the faint end. To quantify this trend, we plot the nucleated fraction as a function of magnitude in the top panel of Figure \ref{nufrac}. The red line with error bars shows our measurements, while the black line reproduces the Virgo cluster core-region ($2\arcdeg \times 2\arcdeg$, total area $\sim 4~\text{deg}^2$) result from \citet{Sanchez19}. The two measurements are in good agreement, considering the relatively large error bar associated with our measurement.

Figure \ref{nuclass} presents the cluster-wide distribution of nucleated and non-nucleated dEs, revealing distinct spatial patterns. Nucleated dEs are concentrated in the region between M87 and M86 (the core of the Virgo cluster), while the density peak of non-nucleated dEs lies north of M87. 
Additionally, nucleated dEs are more strongly clustered around the four main subgroups (M87, M49, M60, and NGC\,4261) than their non-nucleated counterparts.
In contrast, the large-scale distribution of non-nucleated dEs is significantly more extended than that of nucleated dEs.

A comparison of the spatial distributions of dEs with massive galaxies reveals that nucleated dEs are more tightly correlated with the massive galaxy population than non-nucleated dEs. To quantify this difference, we calculate the Kullback--Leibler (K-L) divergence \citep{kll04} between the spatial distribution of massive galaxies and those of the two dE subpopulations. The divergence score is estimated numerically by integrating the product of one distribution and the logarithm of their ratio, see section \ref{kltest}. The resulting value measures how dissimilar the two KDEs are, with zero indicating identical distributions. The significantly lower K-L divergence for nucleated dEs indicates that they more closely trace the environments of massive galaxies, whereas non-nucleated dEs exhibit a comparatively weaker spatial association with massive galaxies.

\begin{figure}
\includegraphics[width=\hsize]{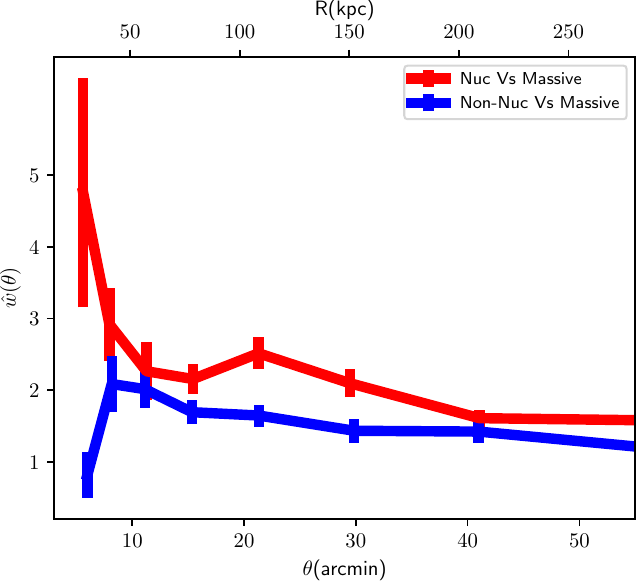}
\caption{
The two-point angular cross-correlation functions are shown for nucleated dEs with massive galaxies (red) and non-nucleated dEs with massive galaxies (blue). Error bars are derived from Jackknifed sub-samples.}
\label{tcorr}
\end{figure}

We additionally analyzed the spatial distribution of dEs relative to massive galaxies using two-point correlation functions \citep{Landy93}. Pair counts were computed with TreeCorr, a Python package optimized for correlation function estimation \citep{Jarvis04}. The angular correlation function, $\hat{\omega}(\theta)$, was measured using 1,000 bootstrap resamplings to assess statistical uncertainties. Figure \ref{tcorr} shows the cross-correlation of dEs with massive galaxies. We find that nucleated dEs exhibit a significantly stronger cross-correlation with massive galaxies than non-nucleated dEs, indicating that nucleated systems preferentially trace the environments of massive galaxies. This difference is especially pronounced at projected separations below $\sim$60 kpc.


The contrasting spatial distributions of dE subclasses likely reflect differences in their formation histories and environmental dependencies. Nucleated dEs are strongly concentrated in high-density regions near massive galaxies and exhibit predominantly old stellar populations, consistent with early formation alongside massive systems during cluster assembly \citep{Oh00}. This trend aligns with the environmental dependence of nuclear star cluster (NSC) formation: the NSC occupation fraction rises sharply in dense regions where tidal interactions and dynamical friction drive the infall and merger of globular clusters, while gas inflows induced by encounters promote in-situ star formation \citep{Carlsten22,Paudel23}. These processes naturally explain the elevated nucleation rates observed near cluster centers.

In contrast, non-nucleated dEs show a broader distribution that follows the cluster’s global potential, suggesting later accretion and weaker exposure to the conditions conducive to NSC formation. Their flatter, more elongated morphologies \citep{Janssen19a, Poulain21} imply lower central binding energies, making them less able to retain gas or star clusters and more prone to tidal disruption near massive galaxies \citep{Paudel14}. Their population may therefore decline in cluster cores over time, weakening any spatial association with massive galaxies.

The concentration of nucleated dEs near the cluster center may also relate to the distribution of ultra-compact dwarfs (UCDs), which are thought to be the stripped remnants of nucleated dwarfs whose outer envelopes were removed by tidal interactions \citep{Goerdt08,Paudel10,Paudel23,Wang23}. This evolutionary link suggests that both nucleated dEs and UCDs trace regions where dynamical processing is most efficient. Overall, the spatial segregation of dE subclasses reflects the combined influence of accretion history, morphological transformation, and the environmentally driven formation and survival of nuclear star clusters.

\newpage

\begin{acknowledgments}

SP and SJY acknowledge support from the Mid-career Researcher Program (RS-2023-00208957 and RS-2024-00344283, respectively) through Korea's National Research Foundation (NRF). 
SJY and CGS acknowledge support from the Basic Science Research Program (2022R1A6A1A03053472 and 2018R1A6A1A06024977, respectively) through Korea's NRF funded by the Ministry of Education. J.Y. was supported by a KIAS Individual Grant (QP089902) via the Quantum Universe Center at Korea Institute for Advanced Study

The DESI Legacy Imaging Surveys consist of three individual and complementary projects: the Dark Energy Camera Legacy Survey (DECaLS), the Beijing-Arizona Sky Survey (BASS), and the Mayall z-band Legacy Survey (MzLS). DECaLS, BASS and MzLS together include data obtained, respectively, at the Blanco telescope, Cerro Tololo Inter-American Observatory, NSF’s NOIRLab; the Bok telescope, Steward Observatory, University of Arizona; and the Mayall telescope, Kitt Peak National Observatory, NOIRLab. NOIRLab is operated by the Association of Universities for Research in Astronomy (AURA) under a cooperative agreement with the National Science Foundation. Pipeline processing and analyses of the data were supported by NOIRLab and the Lawrence Berkeley National Laboratory (LBNL). Legacy Surveys also uses data products from the Near-Earth Object Wide-field Infrared Survey Explorer (NEOWISE), a project of the Jet Propulsion Laboratory/California Institute of Technology, funded by the National Aeronautics and Space Administration. Legacy Surveys was supported by: the Director, Office of Science, Office of High Energy Physics of the U.S. Department of Energy; the National Energy Research Scientific Computing Center, a DOE Office of Science User Facility; the U.S. National Science Foundation, Division of Astronomical Sciences; the National Astronomical Observatories of China, the Chinese Academy of Sciences and the Chinese National Natural Science Foundation. LBNL is managed by the Regents of the University of California under contract to the U.S. Department of Energy. The complete acknowledgments can be found at https://www.legacysurvey.org/acknowledgment/.

This research used data obtained with the Dark Energy Spectroscopic Instrument (DESI). DESI construction and operations is managed by the Lawrence Berkeley National Laboratory. This material is based upon work supported by the U.S. Department of Energy, Office of Science, Office of High-Energy Physics, under Contract No. DE–AC02–05CH11231, and by the National Energy Research Scientific Computing Center, a DOE Office of Science User Facility under the same contract. Additional support for DESI was provided by the U.S. National Science Foundation (NSF), Division of Astronomical Sciences under Contract No. AST-0950945 to the NSF’s National Optical-Infrared Astronomy Research Laboratory; the Science and Technology Facilities Council of the United Kingdom; the Gordon and Betty Moore Foundation; the Heising-Simons Foundation; the French Alternative Energies and Atomic Energy Commission (CEA); the National Council of Science and Technology of Mexico (CONACYT); the Ministry of Science and Innovation of Spain (MICINN), and by the DESI Member Institutions: www.desi.lbl.gov/collaborating-institutions. The DESI collaboration is honored to be permitted to conduct scientific research on Iolkam Du’ag (Kitt Peak), a mountain with particular significance to the Tohono O’odham Nation. Any opinions, findings, and conclusions or recommendations expressed in this material are those of the author(s) and do not necessarily reflect the views of the U.S. National Science Foundation, the U.S. Department of Energy, or any of the listed funding agencies.

\end{acknowledgments}

\begin{contribution}
Sanjaya Paudel and Cristiano G. Sabiu contributed equally. 
\end{contribution}

\facilities{DESI, SDSS, CDS, NED}
\software{
The following software tools were utilized in this work: Astropy \cite{astropy}, Matplotlib \cite{matplotlib}, Numpy \cite{numpy}, Scipy \cite{scipy}, Scikit-learn \cite{Pedregosa11}, Tensorflow \cite{Abadi16}, Seaborn \cite{Waskom21}, TreeCorr  \cite{tree}.
 }


\appendix

\section{Appendix}
\subsection{Comparison of identification}

\begin{figure}[h]
\includegraphics[width=8.4cm]{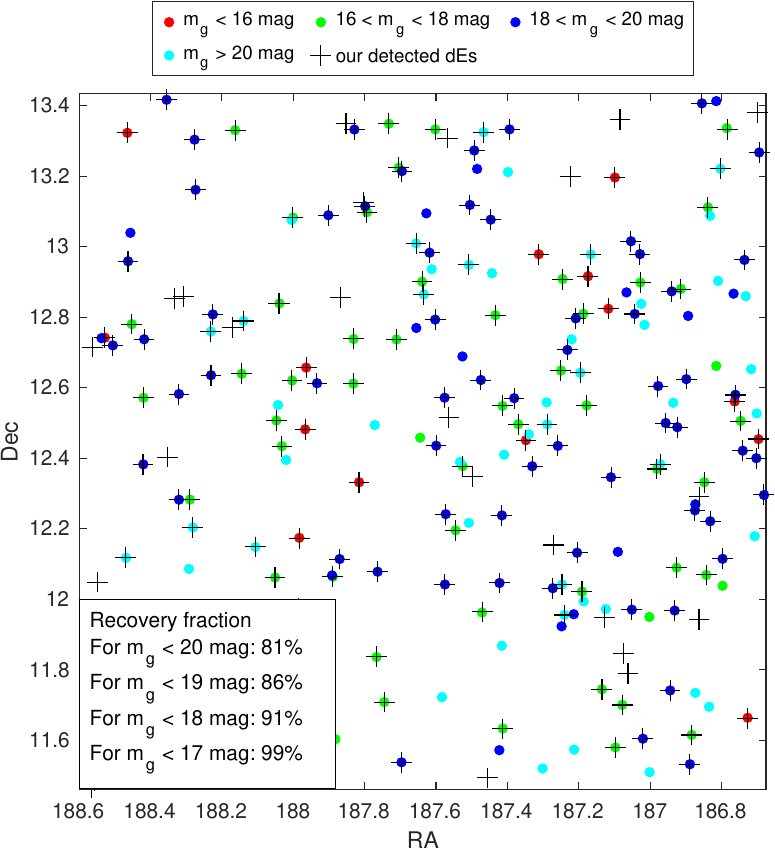}
\caption{
Comparison of our dE identification with that of the NGVS in the core region of the Virgo Cluster. NGVS dEs are represented by filled circles, color-coded by magnitude as indicated in the legend, while our identified dEs are marked with black crosses. Cross marks that do not overlap with filled circles represent additional dEs that were not identified by the NGVS. The inset shows the recovery fraction of dEs as a function of brightness, illustrating detection efficiency across different luminosities.}
\label{recovery}
\end{figure}

\begin{figure}[h]
\includegraphics[width=\hsize]{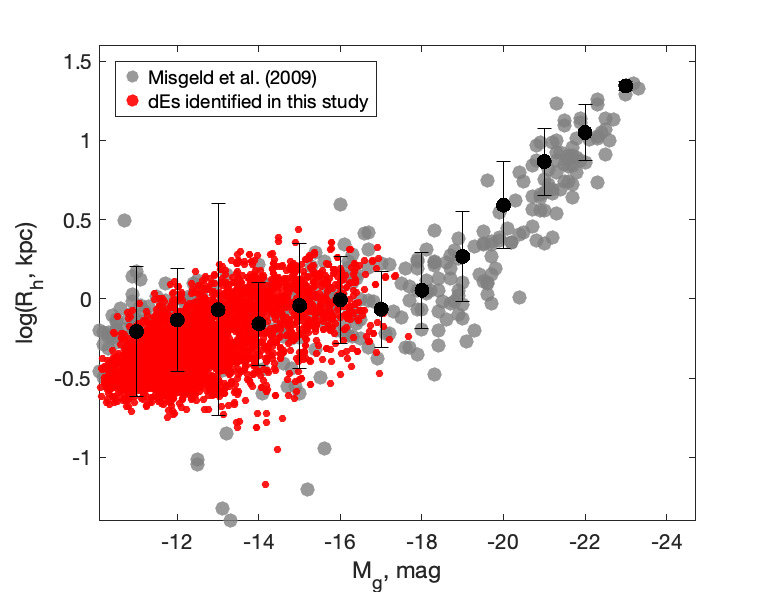}
\caption{
The luminosity$-$size relation of early-type galaxies, including both dEs and massive early-type galaxies. Our identified Virgo Cluster dEs are shown in red, while a reference sample from \cite{Misgeld11} is displayed in light grey. The black error bar represents the 1$\sigma$ scatter of the reference sample in one-magnitude bins.}
\label{scale}
\end{figure}

\begin{figure}
\includegraphics[width=\hsize]{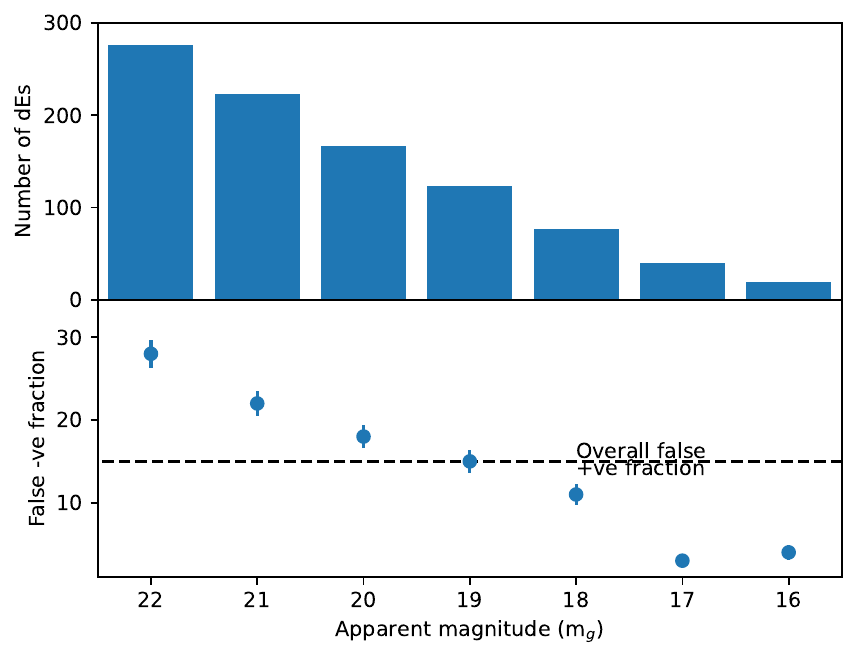}
\caption{
The commutative recovery fraction (100 $-$ False -ve) of our dE detection model compared to the NGVS catalog in the core region of the Virgo Cluster. For dEs brighter than 
 m$_{g}$ $<$20 mag, our model achieves a recovery fraction of 81\%, with an overall false positive rate of 15\%. The upper panel displays the cumulative distribution of dEs as a function of magnitude used in this comparison.}
\label{recv}
\end{figure}

In Figure \ref{recovery}, we compare our dE identification results with those of the Next Generation Virgo Cluster Survey (NGVS), a deep, high-quality imaging project covering the central 100 square degrees of the Virgo Cluster \citep{Ferrarese12}. 
Although the NGVS provides extensive coverage of this region, its detailed catalog focuses on the central 4 square degrees \citep{Ferrarese16}.
Our analysis demonstrates that our semi-automated identification method, which leverages the shallower archival imaging data from the Legacy Survey, successfully recovers nearly 81\% of dEs brighter than 20 mag listed in the NGVS catalog, see Figure \ref{recv}. 
Remarkably, for brighter dEs (with m$_{g}$ $<$ 17 mag), our identification rate reaches 100\%. 
This high recovery rate highlights the robustness of our method for detecting and classifying dEs, even when utilizing large-scale imaging survey data.

Compared to other deep learning image classifiers, EANet achieved superior performance (see Table \ref{modsum}). Among the models evaluated, EANet demonstrates superior performance, achieving a recovery fraction (i.e., 100 - false -ve rate) of up to 81\%, while maintaining a false +ve rate of only 15\%. These results are based on the comparison with a previously well-studied catalog of dwarf galaxies in the central 4 square degrees (2 $\times$ 2) of the NGVS. 
\begin{table}
\centering
\caption{Performance comparison of various deep learning image classification models ($^{*}$ Pre-trainned model).}
\label{modsum}
\scriptsize
\begin{tabular}{c|cccc}
\hline
Model & False -ve & False +v & No. of Par. & Reference \\  
Name & (\%) & (\%) & & \\  
\hline
CNN   & 30 & 40 & 2,622,114 &   \citep{filippov_image_2024}\\
MobileVnet2$^{*}$  & 18 & 35 & 2,257,984 &\citep{Sandler18} \\  
ResNet50$^{*}$& 30 & 10 & 23,587,712 &  \citep{He15} \\  
EAnet & 19 & 15 &  406,928  & \citep{Guo21}\\    
\hline  
\end{tabular}
\end{table}

Due to their smooth visual appearance and well-defined scaling relationships, such as those between luminosity and size, dEs offer more reliable distance estimates compared to star-forming dwarf galaxies, particularly in the absence of redshift information. 
Our primary training catalog focuses on dEs within a redshift range of z $<$ 0.01, consistent with expectations for identifying dEs in the Virgo Cluster.
To validate the identified dEs, we evaluated their conformity to the luminosity$–$size relation, assuming a Virgo Cluster distance of 16.5 Mpc. 
We then performed aperture photometry to measure their sizes and magnitudes, following the methodology described in \cite{Paudel23}. 
Figure \ref{scale} illustrates the luminosity$–$size scaling relation using a well-studied reference sample of early-type galaxies with confirmed distance measurements from \citep{Misgeld11}. 
The identified dEs closely follow this established scaling relation, falling within one standard deviation of the reference sample, further validating their identification as dEs in the Virgo Cluster.

\subsection{Calculation of K-L divergence score}\label{kltest}
To quantify how closely nucleated or non-nucleated dEs follow the spatial distribution of massive galaxies, we use the Kullback--Leibler (K-L) divergence, treating the massive galaxy distribution as a reference $Q(r)$ and the dE distribution as $P(r)$. The K-L divergence is computed as

\[
D_{\rm KL}(P \| Q) = \sum_i P(r_i) \log \frac{P(r_i)}{Q(r_i)}.
\]

Larger values of $D_{\rm KL}$ indicate greater divergence, implying that dEs are more scattered or less tightly correlated with the locations of massive galaxies.

\end{document}